\documentclass[aps,twocolumn,groupedaddress]{revtex4}
\usepackage{graphicx}  
\usepackage{dcolumn}   
\usepackage{bm}        
\usepackage{amssymb}   
\usepackage{multirow}
\usepackage{endnotes}
\usepackage{amsmath, amsthm}
\usepackage[all]{xy}
\usepackage{verbatim}
\usepackage{color}
\usepackage{array,hhline,dcolumn} 

\def\De{\Delta}

\def\to{\rightarrow}

\def\beq{\begin{eqnarray}}
\def\eeq{\end{eqnarray}}

\def\nue{\nu_e}
\def\nueb{\bar{\nu}_e}
\def\numu{\nu_\mu}
\def\numub{\bar{\nu}_\mu}

\def\elp{e^{+}}
\def\elm{e^{-}}
\def\mup{\mu^{+}}
\def\mum{\mu^{-}}
\def\piz{\pi^{0}}

\def\pip{\pi^{+}}
\def\pim{\pi^{-}}
\def\thetmu{\theta_{\mu}}

\def\MBosc1POT{5.58\times 10^{20}}

\begin{document}


\title{Measurement of the neutrino component of an anti-neutrino beam observed by a non-magnetized detector}
\date{\today}
\author{
       A.~A. Aguilar-Arevalo$^{12}$, C.~E.~Anderson$^{15}$, S.~J.~Brice$^{6}$,
       B.~C.~Brown$^{6}$, L.~Bugel$^{11}$, J.~M.~Conrad$^{11}$,
       R.~Dharmapalan$^{1}$, Z.~Djurcic$^{2}$, B.~T.~Fleming$^{15}$, R.~Ford$^{6}$,
       F.~G.~Garcia$^{6}$, G.~T.~Garvey$^{9}$,
       J.~Grange$^{7}$, J.~A.~Green$^{8,9}$, R.~Imlay$^{10}$,
       R.~A. ~Johnson$^{3}$, G.~Karagiorgi$^{11}$, T.~Katori$^{8,11}$,
       T.~Kobilarcik$^{6}$, S.~K.~Linden$^{15}$,
       W.~C.~Louis$^{9}$, K.~B.~M.~Mahn$^{5}$, W.~Marsh$^{6}$,
       C.~Mauger$^{9}$,
       W.~Metcalf$^{10}$, G.~B.~Mills$^{9}$, J.~Mirabal$^{9}$,
       C.~D.~Moore$^{6}$, J.~Mousseau$^{7}$, R.~H.~Nelson$^{4}$,
       V.~Nguyen$^{11}$, P.~Nienaber$^{14}$, J.~A.~Nowak$^{10}$,
       B.~Osmanov$^{7}$, A.~Patch$^{9,11}$, Z.~Pavlovic$^{9}$, D.~Perevalov$^{1}$,
       C.~C.~Polly$^{6}$, H.~Ray$^{7}$, B.~P.~Roe$^{13}$,
       A.~D.~Russell$^{6}$, M.~H.~Shaevitz$^{5}$,
M.~Sorel$^{5}$\footnote{Present address: IFIC, Universidad de
Valencia and CSIC, Valencia 46071, Spain},
       J.~Spitz$^{15}$, I.~Stancu$^{1}$, R.~J.~Stefanski$^{6}$,
       R.~Tayloe$^{8}$, M.~Tzanov$^{4}$, R.~G.~Van~de~Water$^{9}$,
       M.~O.~Wascko$^{10}$\footnote{Present address: Imperial College;
London SW7 2AZ, United Kingdom},
       D.~H.~White$^{9}$, M.~J.~Wilking$^{4}$, G.~P.~Zeller$^{6}$,
       E.~D.~Zimmerman$^{4}$ \\
\smallskip
(The MiniBooNE Collaboration)
\smallskip
}
\smallskip
\smallskip
\affiliation{
$^1$University of Alabama; Tuscaloosa, AL 35487 \\
$^2$Argonne National Laboratory; Argonne, IL 60439 \\
$^3$University of Cincinnati; Cincinnati, OH 45221\\
$^4$University of Colorado; Boulder, CO 80309 \\
$^5$Columbia University; New York, NY 10027 \\
$^6$Fermi National Accelerator Laboratory; Batavia, IL 60510 \\
$^7$University of Florida; Gainesville, FL 32611 \\
$^8$Indiana University; Bloomington, IN 47405 \\
$^9$Los Alamos National Laboratory; Los Alamos, NM 87545 \\
$^{10}$Louisiana State University; Baton Rouge, LA 70803 \\
$^{11}$Massachusetts Institute of Technology; Cambridge, MA 02139 \\
$^{12}$Instituto de Ciencias Nucleares, Universidad Nacional Aut\'onoma de
M\'exico, D.F. 04510, M\'exico \\
$^{13}$University of Michigan; Ann Arbor, MI 48109 \\
$^{14}$Saint Mary's University of Minnesota; Winona, MN 55987 \\
$^{15}$Yale University; New Haven, CT 06520\\
}

\begin{abstract}
Two methods are employed to measure the neutrino flux of the anti-neutrino-mode beam observed by the MiniBooNE detector.  The first method compares data to simulated event rates in a high purity $\numu$ induced charged-current single $\pip$ (CC1$\pip$) sample while the second exploits the difference between the angular distributions of muons created in $\numu$ and $\numub$ charged-current quasi-elastic (CCQE) interactions.  The results from both analyses indicate the prediction of the neutrino flux component of the pre-dominately anti-neutrino beam is over-estimated - the CC1$\pip$ analysis indicates the predicted $\numu$ flux should be scaled by $0.76 \pm 0.11$, while the CCQE angular fit yields $0.65 \pm 0.23$.  The energy spectrum of the flux prediction is checked by repeating the analyses in bins of reconstructed neutrino energy, and the results show that the spectral shape is well modeled.  These analyses are a demonstration of techniques for measuring the neutrino contamination of anti-neutrino beams observed by future non-magnetized detectors.
\end{abstract}

\maketitle
  
\section{\label{sec:Intro} Introduction}  


\indent If $\theta_{13}$ is non-zero, next generation neutrino oscillation experiments will embark on a program to measure the neutrino mass ordering and look for evidence of CP violation in the neutrino sector.  This effort will require precise oscillation measurements with both neutrino and anti-neutrino beams in order to isolate these effects. Since beams produced in an accelerator environment are never purely neutrino nor anti-neutrino in content, detectors must be able to separate the two contributions. Most commonly, this is achieved by employing a magnetic field to identify the final-state $\mu^{-}$ (or $\mu^+$) produced in charged-current $\nu_{\mu}$ (or $\bar{\nu}_{\mu}$) interactions. A handle on the overall level and energy dependence of $\nu_{\mu}$ versus $\bar{\nu}_\mu$ induced events, however, is also possible in unmagnetized detectors with a suitable choice of reaction channels.

\indent Accelerator-based neutrino beams are typically created by colliding proton beams with thick nuclear targets.  Mesons produced at a variety of energies and angles are focused by a magnetic horn before entering a decay tunnel.  Meson decays can be calculated sufficiently well for a given beam geometry that the neutrino flux uncertainties arise mainly from uncertainties in the meson production cross sections. In particular, to avoid extrapolating data taken with diverse nuclear target materials or proton energies, neutrino experiments require dedicated hadron production cross-section measurements taken with the same beam energy and target to obtain a reliable flux prediction. If an accelerator-based neutrino experiment lacks such hadron production data, it may be able to meet its oscillation analysis goals using calibrations from a near detector; however the secondary physics goal of measuring neutrino-nucleon absolute cross sections will still be limited by flux uncertainties.

\indent To avoid ambiguity, in this paper references to ``neutrinos" are not meant to also refer to anti-neutrinos, and ``mode" refers to the polarity of the magnetic focusing horn used in the beamline.  In this way, for example, ``anti-neutrino events" refers to anti-neutrino induced events exclusively while ``anti-neutrino-mode events" refers to data obtained when the horn polarity focuses negatively charged particles, which is a mix of neutrino and anti-neutrino induced events.

\begin{figure}[h]
\begin{center}$
\begin{array}{c}
\includegraphics[scale=0.51]{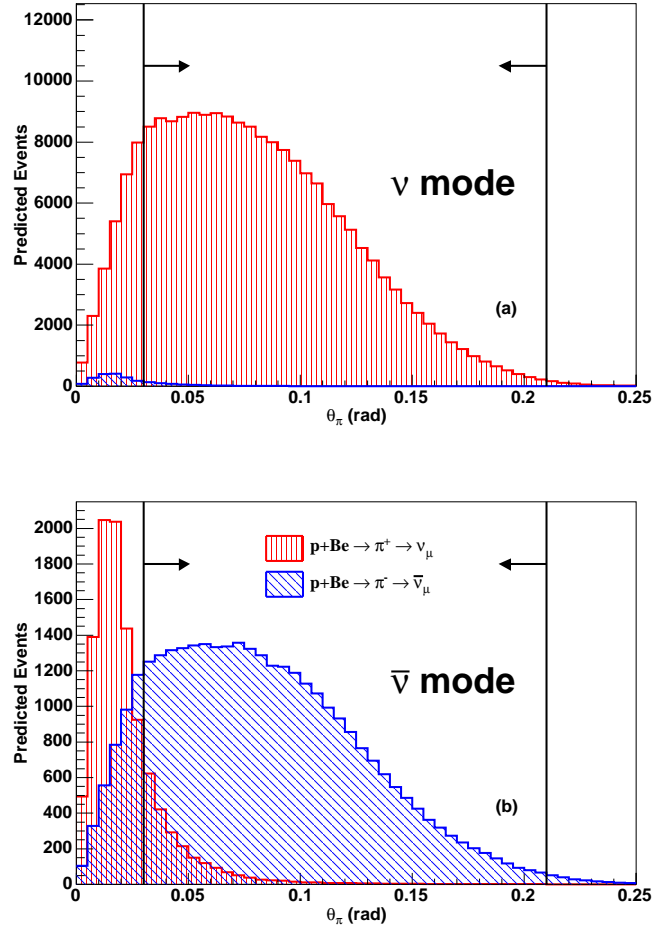} \\
\end{array}$
\end{center}
\caption{(Color online) Predicted angular distributions of pions with respect to the incident proton beam ($\theta_{\pi}$) producing $\numu$ and $\numub$ in (a) neutrino mode and (b) anti-neutrino mode.  Only pions leading to $\numu$ and $\numub$ events in the detector are shown, and all distributions are normalized to 5.66 $\times \,\, 10^{20}$ protons on target.  Arrows indicate the region where HARP data\cite{HARP} are available.}
\label{fig:pionAngles}
\end{figure}

\indent The Mini Booster Neutrino Experiment (MiniBooNE) is located at Fermilab in Batavia, Illinois and has made many oscillation~\cite{osc1,mbLowE,prlDisap,nubOsc1,nubOsc2} and cross section~\cite{MB_PRL,piQEratio,NCpi0,qePRD,NCE,CCpi0,CCpip,cohPi0} measurements.  For MiniBooNE, the pion production data crucial to the flux model come from proton-beryllium cross sections on a 5\% interaction length target reported by the HARP experiment~\cite{HARP}.  However, even with dedicated data appropriate to the experimental setup of MiniBooNE, there remain small regions of phase space relevant to MiniBooNE not covered by these HARP measurements.  Of particular importance is the production of very forward pions with respect to the direction of the incoming proton beam.  This is the dominant production region of parent particles contributing neutrinos to the anti-neutrino-mode beam, or vice versa.  MiniBooNE uses a magnetic horn to defocus the majority of these background parent particles, but as Figure~\ref{fig:pionAngles} suggests, the very forward pions can escape magnetic deflection.  This same angular region suffers from a sizable beam-related proton background and would also require a model-dependent acceptance correction~\cite{HARP30mrad}.  For these reasons, pion cross sections in the $\theta_{\pi}$ \textless\,30~mrad region, where $\theta_{\pi}$ is the angle the outgoing pion makes with respect to the incoming proton beam, are not reported by HARP and the majority of the MiniBooNE flux prediction arising from $\pip$ ($\pim$) decay while focusing $\pim$ ($\pip$) is extrapolated from the available hadron production data.  The hadron production data cover $\sim$~90\% of sign selected pions, while less than 25\% of oppositely charged pions in the same beam are constrained.  Some of these acceptance limitations could be reduced by use of the long-target data taken by HARP, which are actively being analyzed.

\indent The overall contamination rate is more significant in anti-neutrino mode due to effects from both flux and cross section: the leading-particle effect at the target preferentially produces about twice as many $\pip$ as $\pim$, and the neutrino cross section is about three times higher than  the anti-neutrino cross section in the MiniBooNE energy range ($\sim$ 1~GeV)~\cite{nuance}.  For these reasons anti-neutrino induced events are not a serious complication for neutrino-mode running, as these flux and cross-section effects conspire to suppress their contribution, while the same effects amplify the neutrino contamination in anti-neutrino-mode data.  Simulation predicts anti-neutrino events account for $\sim 1\%$ of neutrino-mode data while neutrinos are responsible for $\sim 30\%$ of anti-neutrino-mode data.  This motivates a dedicated study of the neutrino flux contribution to the anti-neutrino-mode data.  A data set corresponding to 5.66 $\times 10^{20}$ protons on target is analyzed here and is important for both the on-going MiniBooNE anti-neutrino oscillation search~\cite{nubOsc1,nubOsc2} and anti-neutrino cross-section measurements.


\indent Two approaches for measuring the neutrino flux in anti-neutrino-mode are taken.  In the first method, a high purity sample of charged-current single $\pip$ (CC1$\pip$) events isolate the $\numu$ contribution in the beam.  The second method exploits the interference term in the charged-current quasi-elastic (CCQE) cross section, where the angular distribution of final-state muons are predicted to be distinct for $\numu$ compared to $\numub$ interactions.  Both techniques were introduced in the MiniBooNE anti-neutrino-mode run proposal~\cite{LOI}.  These two approaches offer complementary means of measuring the neutrino flux component in anti-neutrino-mode data, with the CCQE sample providing a constraint at lower neutrino energies while the CC1$\pi^+$ measurement covers higher energies. They provide both a check of the MiniBooNE beam simulation in a region not covered by external data and demonstrate a set of techniques for measuring the $\numu$ contamination in an anti-neutrino-mode beam in the absence of a magnetized detector.  It has been argued elsewhere that even modest statistical separation of charged-current neutrino and anti-neutrino events, afforded by the kind of analyses presented here, may be sufficient to meet the physics goals in proposed future experiments such as neutrino factories~\cite{HubSch}.

\indent This paper is organized as follows: the MiniBooNE experiment is described in Section~\ref{sec:BooNE} while Section~\ref{sec:nuInt} details the neutrino and anti-neutrino scattering models.  Two techniques to measure the neutrino contribution to the anti-neutrino flux are presented in Sections~\ref{sec:CC1pi} and~\ref{QE}.  The results are compared in Section~\ref{sec:comp}, implications for other neutrino experiments are discussed in Section~\ref{sec:imps} and this work is summarized in Section~\ref{sec:conc}.

\section{\label{sec:BooNE} The MiniBooNE Experiment}


\subsection{\label{sbsec:mbFlux} Beamline and flux}

\indent The Booster Neutrino Beamline (BNB) provides the neutrino and anti-neutrino flux to MiniBooNE. A beam of 8~GeV kinetic energy protons is extracted from the Booster synchrotron in ``spills" of $5 \times 10^{12}$ protons over 1.6~$\mu$s at a maximum rate of 5 Hz.  A lattice of alternatively focusing and defocusing quadrupole magnets steer the proton bunches to a beryllium target 71 cm (1.75 interaction lengths) long.  The protons collide with the target to create a spray of secondary particles.  An aluminum electromagnetic horn surrounding the target is pulsed to coincide with the p-Be collisions, creating a toroidal magnetic field to focus mesons of the desired charge.  The horn pulses are such that the magnetic field is constant for the duration of the proton spill.  In neutrino mode, the magnetic horn focuses positively charged secondary particles while defocusing those with negative charge; the horn effects are reversed in anti-neutrino mode.  The focused mesons are allowed to decay in a 50~m air-filled decay region which terminates at a steel beam dump.  The dominant decay modes of the mesons, mostly pions, produce muon neutrinos and anti-neutrinos.  

\begin{figure}[h]
\begin{center}$
\begin{array}{c}
\includegraphics[scale=0.51]{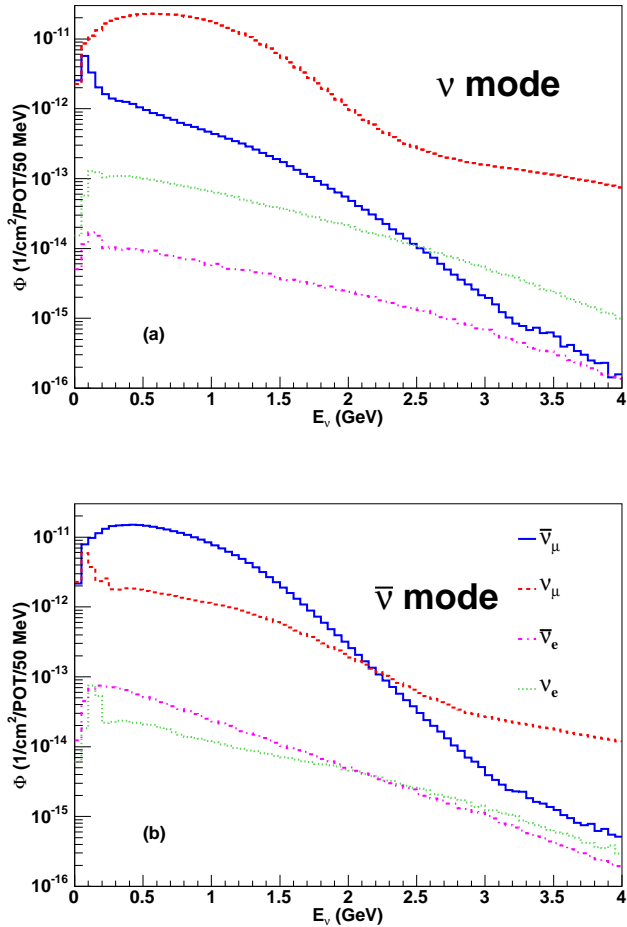} \\
\end{array}$
\end{center}
\caption{(Color online) The MiniBooNE flux prediction for (a) neutrino mode and (b) anti-neutrino mode.  
Due to the leading-particle effect, the neutrino contribution to the anti-neutrino-mode flux is more significant compared to the anti-neutrino component of the neutrino-mode beam.  Plots taken from Ref.~\cite{mbFlux}.}
\label{fig:fluxes}
\end{figure}

\indent A \textsc{geant}4-based model~\cite{GEANT4} is used to predict the neutrino and anti-neutrino flux at the detector.  The simulation considers proton travel to the target, p-Be interactions in the target including meson production, magnetic horn focusing, particle propagation, meson decay, and finally neutrino and anti-neutrino transport to the detector.  As mentioned earlier, measurement of pion cross sections from p-Be interactions are obtained from the HARP experiment.  The HARP double differential cross-section error matrix is used to set pion production uncertainties~\cite{mbFlux}.  Even with valuable data constraints, meson production at the target contributes the largest systematic error to the flux prediction.  The fractional uncertainty on pion production is $\sim 8\%$ around the flux peak, while the uncertainty grows significantly in regions dominated by pions unconstrained by HARP data.  The flux prediction in neutrino and anti-neutrino modes is presented in Figure~\ref{fig:fluxes}.  Details of the beamline and flux prediction are given in Ref.~\cite{mbFlux}.

\subsection{\label{sbsec:mbDet} Detector}

The MiniBooNE detector is a 6.1 m radius sphere filled with 818 tons of pure Marcol7 mineral oil.  It houses 1520 8-inch Hamamatsu photomultiplier tubes (PMTs) segregated into two optically isolated regions: an inner signal region of 575 cm radius and an outer veto shell of thickness 35 cm.  The former contains 1280 PMTs (11.3\% coverage) while the latter holds 240 PMTs.  The veto region is used to enforce containment of charged particles produced by neutrinos and anti-neutrinos from the beam and reject charged particles entering the tank. 

\indent The mineral oil has a density of 0.845 g/cm$^{3}$ with an index of refraction of 1.47 at $20\,^{\circ}\,\mathrm{C}$.  Charged particles with velocity $\beta\, \textgreater\, 0.68$ produce Cherenkov radiation.  Particle identification and reconstruction is principally obtained through the pattern and timing of this prompt Cherenkov light; however, delayed scintillation light present due to fluorescent components in the oil has also been used effectively to provide energy information for charged particles produced below Cherenkov threshold~\cite{NCE}.  

\indent MiniBooNE electronics record PMT charge and time information beginning about 5~$\mu$s before the 1.6~$\mu$s BNB proton delivery.  Data are recorded for a total of 19.2~$\mu$s.  The 5 $\mu$s interval before the beam spill is primarily present to minimize data contamination caused by cosmic ray muons stopping in the signal region prior to the start of the DAQ window.  PMT activity is recorded for more than 10~$\mu$s after beam delivery to observe electrons from the at-rest decay of muons (hereafter referred to as ``Michel" electrons) subsequent to the initial neutrino or anti-neutrino induced interaction.  

\indent The detector response to muons is measured using a dedicated muon tagging system that independently measures the energy and direction of cosmic ray muons up to 800 MeV.  MiniBooNE employs a scintillator hodoscope directly above the detector and seven internal scintillator cubes at different depths, each connected to a dedicated one-inch PMT for readout.  The measured ranges and directions of muons traversing the hodoscope and stopping in cubes are used to verify muon reconstruction algorithms.  
The energy (angle) resolution improves from 12\% (5.4~deg) at 100~MeV to 3.4\% (1.0 deg) at 800 MeV.  Full detector details and calibrations are available in Ref.~\cite{mbDet}.

\subsection{\label{sbsec:detSim} Detector simulation}

\indent The detector response to particle interactions and propagation is simulated using \textsc{geant}3~\cite{geant3}.  The entire detector geometry is considered, including the steel tank, external supports and main inner components.  In addition, the surrounding environment composed of dirt external to the MiniBooNE enclosure, the concrete cylindrical housing and the air-filled gap between the detector and walls is treated.  Of critical importance is the treatment of particle transport in the detector medium.  The \textsc{geant}3 program takes as input the final-state particles emerging from the nucleus and simulates their propagation in the detector. 


\indent With a few exceptions, MiniBooNE uses the standard \textsc{geant}3 settings to simulate physics processes.  Deviations include a custom model for light propagation in the detector oil and a substitution of the hadronic interaction model.  The default \textsc{gfluka} hadron model is replaced by the \textsc{gcalor}~\cite{gcalor} package, which better models pion absorption ($\pi^{\pm} + X \to X^{'}$) and charge exchange ($\pi^{\pm} + X \leftrightarrow \piz + X^{'}$) processes.  This is particularly relevant for the present analysis, where the predicted event composition of the two interaction samples studied is dependent on the pion survival model.  Based on comparisons with external data~\cite{piAbsCEXdata} and the \textsc{gcalor} prediction, an uncertainty of 35\% (50\%) is assigned to the pion absorption (charge exchange) interaction \emph{in the detector medium}.  The uncertainty for the same processes \emph{inside the nucleus} is discussed in Section~\ref{sbsec:finState}.

\indent The model for light propagation in the oil is formed using a combination of external measurements and calibration data.  Photon emission through Cherenkov and scintillation processes is simulated and propagated until the photon either is absorbed or hits a PMT photocathode, possibly leading to photoelectron production.  Light emmission, attenuation and scattering are included.  The optical model of the detector describes the wavelength, time, and angular dependence of these processes~\cite{OM}.   


\section{\label{sec:nuInt} Predicted neutrino and anti-neutrino interactions}

MiniBooNE uses the \textsc{nuance}~\cite{nuance} event generator to simulate neutrino and anti-neutrino interactions in the detector.  \textsc{nuance} includes a comprehensive neutrino and anti-neutrino cross section model which considers known interactions in the neutrino and anti-neutrino energy range from $\sim$ 100~MeV to 1~TeV.  Ninety-nine reactions are modeled separately and combined with nuclear models describing bound nucleon states and final-state interactions to predict event rates and kinematics.

Bound nucleons in the detector medium are described by the Relativistic Fermi Gas model~\cite{RFG}.  This assumes the nucleons to be independent and quasi-free.  Also specified is a hard cut-off in available struck nucleon energies as dictated by the exclusion principle.  

The neutrino and anti-neutrino interaction types relevant to the analysis presented here are charged-current quasi-elastic (Section~\ref{sbsec:QExsec}) and pion production (Section~\ref{sbsec:firstPi}).  The neutrino-induced absolute cross sections for both processes have been measured at MiniBooNE using a flux prediction well determined by HARP data.  These cross-section measurements are utilized in the anti-neutrino-mode simulation.

\subsection{\label{sbsec:QExsec} Charged current quasi-elastic scattering}

To model CCQE interactions, this analysis uses measured cross sections
from the MiniBooNE neutrino mode CCQE data~\cite{qePRD} and a model which has
been found to well-reproduce the kinematics of such events. Specifically,
MiniBooNE adopts the CCQE scattering formalism of Smith-Moniz~\cite{RFG}.
The vector component of the interaction is measured by electron scattering
experiments and is assumed to have a non-dipole form~\cite{BBA}. The axial-vector
form factor employs a dipole construction, containing an ``axial mass", $M_{A}$,
taken either from MiniBooNE or external data, depending on the neutrino target.

The MiniBooNE mineral oil is composed of C$_{n}$H$_{2n+2}$, $n \sim 20$, and the prediction for CCQE
scattering is different for the two flavors of target. In the present analysis, $M_{A}^{\mathrm{eff}}$ = 1.35 $\pm$ 0.17~GeV together with a Pauli
blocking adjustment, $\kappa$ =1.007 $\pm$ 0.012 are assumed for bound
nucleon scattering. These values come from a high statistics analysis of
MiniBooNE $\numu$ CCQE events on carbon~\cite{qePRD} and are consistent with
values recently determined from an independent MiniBooNE neutral-current
elastic scattering sample~\cite{NCE}. A previous shape-only study has shown
that these CCQE model parameters reproduce the MiniBooNE anti-neutrino-mode data shape~\cite{JG_nuint}, and 
therefore the same $M_{A}^{\mathrm{eff}}$ and $\kappa$ values are
applied to both $\numu$ and $\numub$ CCQE scattering events on carbon.

For free scattering off hydrogen, a process accessible to anti-neutrino and
not neutrino CCQE events, a value of $M_{A}$ = 1.03 $\pm$ 0.02~GeV is used based
on a global fit to previous light target data~\cite{MAMeas}.

In the case of carbon scattering, the superscript "eff", short for ``effective",
on $M_{A}$ is introduced to allow for the possibility that nuclear effects are
responsible for the apparent discrepancy between the MiniBooNE carbon-based
measurements and light target results. This is also theoretically motivated by
a possible reconciliation between the measurements through a mechanism
resulting in intranuclear correlations of greater importance than previously
thought~\cite{Martini1,Martini2,Nieves,Nieves2,Amaro}. Such a mechanism would indicate a larger CCQE cross section
for nuclear targets than for free scattering, which in this case, is reflected in
the higher $M_{A}$ choice for carbon versus hydrogen scattering.

\subsection{\label{sbsec:firstPi} Pion production}

Baryonic resonances are the dominant source of single pion production at MiniBooNE.  The formalism to describe these events is taken from the Rein-Sehgal model~\cite{R-S}, where the relativistic harmonic oscillator quark model is assumed~\cite{Feyn}.  Eighteen resonances are considered, however the $\De$(1232) is dominant in the energy range spanned by MiniBooNE.  Multi-pion production mechanisms are also considered, though their contribution is predicted to be small.

The axial masses in the resonance channels are set simultaneously to reproduce inclusive non-MiniBooNE charged-current data~\cite{K2Kpr}.  The extracted values are $M_{A}^{1\pi} = 1.10 \pm 0.27$~GeV (single pion production) and $M_{A}^{\mathrm{multi}-\pi} = 1.30 \pm 0.52$~GeV (multi-pion production). 

In the present analysis the charged-current single $\pip$ (CC1$\pip$) prediction with these assumptions is adjusted to reproduce the kinematic distributions measured in MiniBooNE neutrino-mode data~\cite{qePRD,CCpip}.


\subsection{\label{sbsec:finState} Final state interactions}

For a neutrino or anti-neutrino interaction with a nucleon bound in carbon, \textsc{nuance} propagates the outgoing hadrons including nucleons, mesons and baryonic resonances, and simulates their re-interaction as they exit the nucleus.  The initial interaction model employs the impulse approximation which assumes an instantaneous exchange with independent nucleons.  Subsequent to the initial neutrino or anti-neutrino interaction, particles produced inside the nucleus are propagated step-wise in 0.3 fm increments until they emerge from the $\sim$ 2.5 fm radius sphere.  Intermittently, the probability for hadronic re-interaction is calculated using a radially-dependent nucleon density distribution~\cite{nuclDist} along with external $\pi - N, N - N$ cross-section measurements~\cite{piN}.  For $\De$ re-interactions ($\De + N \to N + N$), an energy-independent probability of 20\% (10\%) is taken for $\De^{+} + N$, $\De^{0} + N$ ($\De^{++} + N, \De^{-} + N$) based on K2K data~\cite{K2Kpr} and is assigned 100\% uncertainty.  

As mentioned earlier, out of all hadronic re-interaction processes, pion absorption and charge exchange ($\pi^{\pm} + X \leftrightarrow \piz + X^{'}$) are the most relevant in predicting the composition of the CC1$\pi^+$ (Section~\ref{sbsec:3se}) and CCQE (Section~\ref{sec:QEsam}) samples studied in this analysis.  \emph{Intranuclear} fractional uncertainties on pion absorption (charge-exchange) are set to 25\% (30\%) based on comparisons between external data~\cite{piAbsCEXdata} and \textsc{nuance}.  The simulation of these two processes in the detector medium is addressed separately in the detector simulation (Section~\ref{sbsec:detSim}).  

\section{\label{sec:CC1pi} Measuring the Neutrino Flux Component in the CC1$\pip$ Sample}

\subsection{\label{sbsec:3se} The CC1$\pip$ sample}

The events in the CC1$\pip$ sample in anti-neutrino mode originate almost exclusively from $\numu$ interactions, making it an excellent candidate for measuring the $\numu$ content of the anti-neutrino-mode beam.  In the few-GeV energy range, the dominant charged-current single pion production channels contain a final-state $\pip$ ($\pim$) in the case of $\numu$ ($\numub$) scattering.  MiniBooNE cleanly identifies CC1$\pip$ events by selecting 3 ``subevents", attributed to the muon from the primary $\numu$ interaction and two subsequent decay electrons, one each from the $\mum$ and $\pip$ decay chain:

\begin{equation}
\begin{array}{cccl}
\label{eqn:ccpip}
&1:&  \hspace{-1.0in} \numu +p(n)  \to  \mum + p(n) + \pip  \\
&  &          \hspace{1.2in} \hookrightarrow  \mup + \numu \\
&2:&          \hspace{-0.1in} \hookrightarrow \elm + \nueb + \numu  \\
&3:&          \hspace{2.0in} \hookrightarrow \elp + \nue  + \numub. 
\end{array} 
\end{equation}

The mono-energetic $\mup$ from the decay of stopped $\pip$ does not lead to a separate subevent due to the short lifetime of the $\pip$. Subevents are defined as clusters in time of PMT activity (or PMT ``hits").  A hit is any PMT pulse passing the discriminator threshold of $\sim$ 0.1 photoelectrons.  A temporal cluster of PMT activity with at least 10 hits within a 200 ns window and individual hit times less than 10 ns apart, while allowing for at most two spacings of 10 - 20 ns, defines a subevent.  Apart from detection efficiencies, some neutrino-induced CC1$\pip$ events do not enter the three subevent sample as $\sim$ 8\% of $\mum$ are captured in carbon~\cite{mumCap} and therefore do not lead to the production of a Michel electron.  Other selection cuts made to enhance sample purity and improve reconstruction are given with efficiencies in Table~\ref{tbl:EffsCC1pi}.  Cut 1 is the three subevent criterion previously detailed.  Cut 2 requires that the first subevent occur during a 3~$\mu$s time window centered on the BNB proton spill.  Cut 3 rejects events close to the detector edge that are likely to be poorly reconstructed.  Selection cuts on the number of tank hits are based primarily on the observation that Michel electrons produce fewer than 200 tank hits.  Cut 4 ensures the first subevent is not a Michel electron and rejects low energy muons that might be reconstructed poorly.  Cut 5 requires that the number of hits for the second and third subevents is consistent with a Michel electron.  Veto PMT activity is monitored simultaneously with the main tank PMTs, thus Cut 6 ensures no subevent is due to charged particles entering the tank and that all charged particles produced inside the detector are contained.  
\indent Cut 7 enforces spatial correlation between the end of the muon track and the closest Michel electron vertex.  This reduces a class of backgrounds where neither the second nor the third subevent arise from the decay of the muon to a Michel electron.  This cut is applied only to the Michel closest to the end of the reconstructed primary muon track as the pion lifetime compared to the muon is short enough that either Michel can occur temporally first.

Charged-current single $\pim$ events induced by $\numub$ are largely rejected by the primary requirement of three subevents because most $\pim$ come to rest and are captured by carbon nuclei~\cite{pimCap}, yielding no decay electron. 
The predicted event composition after this selection is presented in Table~\ref{tbl:ccpiPur}.  The sample is $82\%$ observable CC1$\pi^+$ events (i.e., events with a single muon, a single $\pip$, and any number of nucleons exiting the initial target nucleus). Some $\numub$ CC1$\pim$ events do make it into the sample, primarily due to decay-in-flight $\pim$. Starting from an event population that is $\sim~70\%$ $\numub$, this simple two decay electron requirement remarkably yields a sample that is $\sim~80\%$ pure $\numu$.

\begin{table}
\caption{\label{tbl:EffsCC1pi} Summary of selection cuts in the CC1$\pip$ sample.  Purity and efficiency numbers are sequential and are calculated for the ``observable CC1$\pip$" event signature - 1 $\mum$, 1 $\pip$.}
\begin{ruledtabular}
\begin{tabular}{cccc}
 \multirow{2}{*}{Cut \#} & \multirow{2}{*}{Description} & Efficiency & Purity \\ 
& & (\%) & (\%) \\
\hline
\vspace{1 mm}
0 & No cuts & 100 & 10 \\
 \vspace{1 mm}
1 & Three subevents & 30 & 29 \\
 \multirow{2}{*}{\,\,2} & 1st subevent in event time window & \multirow{2}{*}{28} & \multirow{2}{*}{34} \\
 \vspace{1 mm}
 & 4000 \textless\, T(ns) \textless\, 7000 & & \\
\multirow{2}{*}{\,\,3} & All subevents: reconstructed & \multirow{2}{*}{23} & \multirow{2}{*}{36} \\
 \vspace{1 mm}
 & vertex \textless\, 500~cm from tank center & & \\
 \vspace{1 mm}
4 & 1st subevent: tank hits \textgreater\, 200 & 22 & 39 \\
\multirow{2}{*}{\,\,5} & 2nd, 3rd subevents:  &  \multirow{2}{*}{19} &  \multirow{2}{*}{65} \\
 \vspace{1 mm}
& tank hits \textless\, 200 &  & \\
 \vspace{1 mm}
6 & All subevents: veto hits \textless \, 6 & 16 & 78 \\
 \multirow{3}{*}{\,\,7} & Distance between reconstructed & \multirow{3}{*}{12} & \multirow{3}{*}{82} \\
& end of 1st subevent and nearest & \\
& Michel electron vertex \textless\, 150~cm & & \\
\end{tabular}
\end{ruledtabular}
\end{table}

\begin{table}[h]
\caption{\label{tbl:ccpiPur} Predicted event composition of the CC1$\pip$ sample in anti-neutrino mode.}
\begin{ruledtabular}
\begin{tabular}{cc}
Interaction Channel  & Contribution (\%) \\
\hline  
$\numu\, N \to \mum \,\pip\, N$ (resonant) & 64 \\
$\numu\, A \to \mum \,\pip\, A$ (coherent) & 7 \\
$\numub\, N \to \mup \,\pim\, N$ (resonant) & 6 \\
$\numu \, n \to \mum\, p$ & 6 \\
$\numu \, n \to \mum \,\piz\, p$ & 2 \\
$\numub \, p \to \mup \, \piz\, n$ & 1 \\
Other (mostly DIS) & 14 \\
\hline
``Observable CC1$\pip$" & \multirow{2}{*}{82} \\
(1 $\mum$, 1 $\pip$) & \\
\end{tabular}
\end{ruledtabular}
\end{table}

\subsection{\label{sbsec:ccpiEvntReco} CC1$\pip$ event reconstruction}

\indent In this analysis, charged-current single $\pip$ event reconstruction relies exclusively on the observation of the outgoing muon.  Muon kinematics are obtained by the pattern, timing, and total charge of prompt Cherenkov radiation collected by PMTs in the first subevent of the interaction.  The topology and timing of the observed PMT hits are compared to a likelihood function operating under a muon hypothesis.  This likelihood function predicts hit patterns and timing based on the interaction vertex and the momentum four-vector of the muon.  The likelihood function simultaneously varies these seven parameters while comparing to the observed PMT hits.  The parameters from the maximized likelihood function yield the reconstructed muon kinematics.  

\begin{figure}[h]
\begin{center}
\includegraphics[scale=0.46]{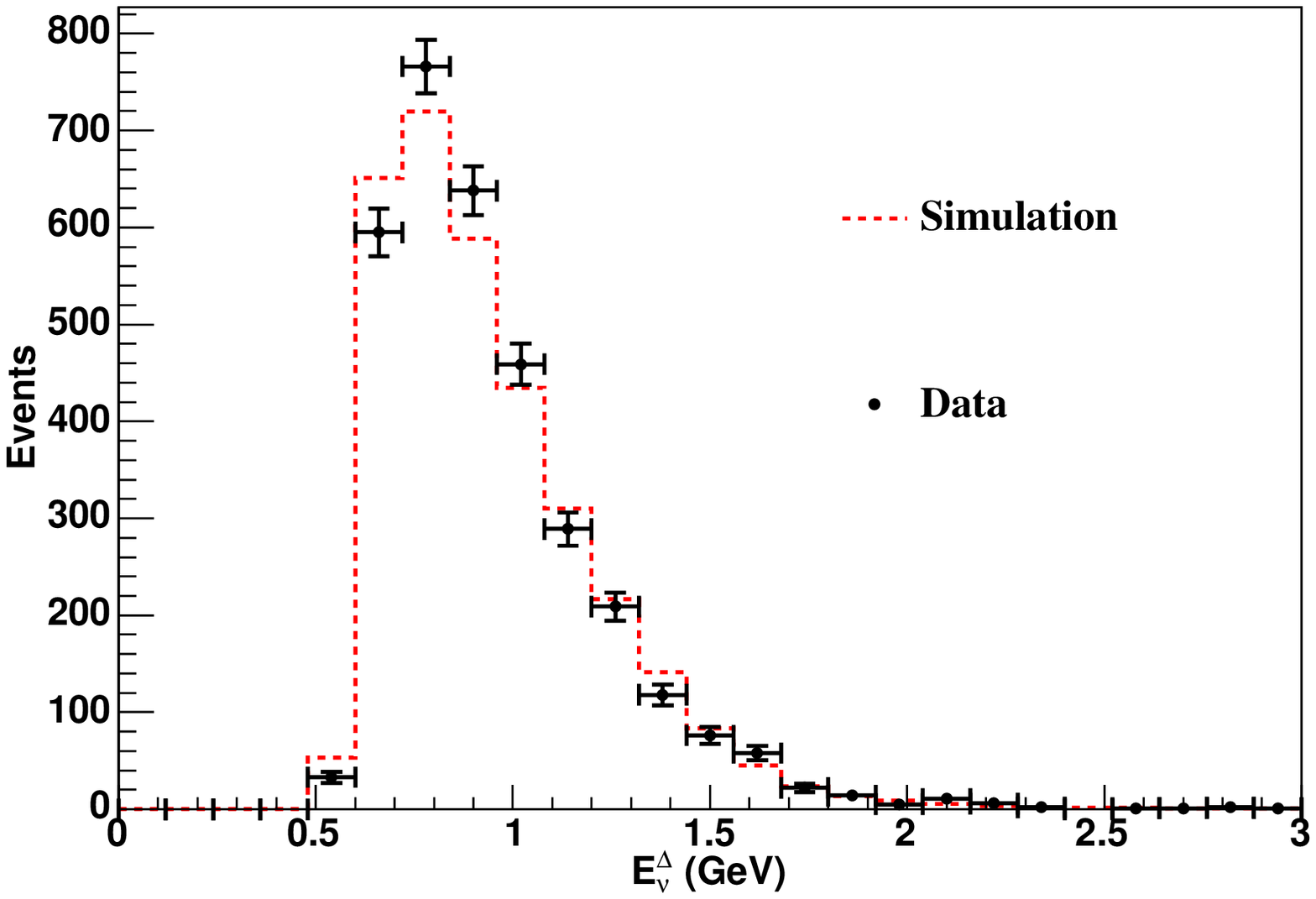}
\caption{(Color online) The reconstructed energy spectrum for simulation versus data in the anti-neutrino-mode CC1$\pip$ sample. Simulation is normalized to data, and only statistical errors are shown.}
\label{fig:EnuPismear}
\end{center}
\end{figure}

\indent Under the assumption of $\De$(1232) production by a neutrino scattering off a stationary nucleon target in carbon, the neutrino energy is given by:

\begin{center}
\begin{equation}
\label{eqn:EnuPi}
E_{\nu}^{\mathrm{\De}}= \frac{2 \left(M_{p} - E_{B}\right) E_{\mu} - \left( E^{2}_{B} - 2 M_{p} E_{B} + m^{2}_{\mu} + \De M^{'2} \right)}{2\left[ \left(M_{p} - E_{B} \right) - E_{\mu} + p_{\mu}\,\text{cos}\,\theta_{\mu}\right]}\end{equation}
\par\end{center}

\noindent where E$_{B} = 34$~MeV is the binding energy, $m_{\mu}$ is the muon mass, $\De M^{'2} = M_{p}^{2} - M_{\De}^{2}$, where $M_{\De}$ ($M_{p}$) is the $\De$(1232) (proton) mass, $p_{\mu}$ is the muon momentum, and $\thetmu$ is the outgoing muon angle relative to the incoming neutrino beam.  Effects not accounted for in the reconstruction include non-resonant pion production, contributions from higher mass $\De$ resonances and scattering off the quasi-free protons in hydrogen instead of carbon.  A shape comparison of reconstructed $E_{\nu}^{\mathrm{\De}}$ in data and simulation is presented in Figure~\ref{fig:EnuPismear}.

\subsection{\label{sbsec:ccpiWSmeas} Measuring the neutrino flux component in the anti-neutrino-mode CC1$\pip$ sample}

The simulation sample is separated into two components: observable CC1$\pip$ events and background.  All observable CC1$\pip$ events in the simulation are modeled using the CC1$\pip$ cross section that has been measured in MiniBooNE neutrino-mode data~\cite{CCpip}.  Given that the majority of the CC1$\pip$ sample in anti-neutrino mode is induced by neutrinos, with this cross-section measurement applied any remaining normalization difference between data and simulation is interpreted as a neutrino flux measurement.  Results are presented in Table~\ref{tbl:wsCCpi}.  Events in the anti-neutrino mode CC1$\pip$ sample indicate the neutrino flux in anti-neutrino mode is lower than the simulation predicts.  The extracted calibration is 0.76 $\pm$ 0.11 of the nominal prediction over all reconstructed energies, while the analysis applied to individual energy ranges does not indicate any significant energy dependence.  

\begin{table}
\caption{\label{tbl:wsCCpi}  Anti-neutrino-mode CC$1\pip$ sample details and $\numu$ flux component measurement.  The measured cross section has been applied to simulation, and the $\numu$ flux scale is found by calculating (observed events - expected $\numub$ events) / expected $\numu$ events.  The reported error is discussed in more detail in Section~\ref{sbsec:cc1piErrs}.  The Monte Carlo sample is generated so that the associated statistical error is negligible compared to the other sources of uncertainty.}
\begin{ruledtabular}
\begin{tabular}{cccccc}

E$_{\nu}^{\mathrm{\De}}$ Range & Mean Gen. & Events & \multicolumn{2}{c}{Expected} & $\numu$ Flux \\
(MeV) & E$_{\nu}$ (MeV) & in Data & $\numu$ & $\numub$ & Scale \\
\hline  
600 - 700 & 961 & 465 & 556 & 104 & 0.65 $\pm$ 0.10 \\
700 - 800 & 1072 & 643 & 666 & 118 & 0.79 $\pm$ 0.10 \\
800 - 900 & 1181 & 573 & 586 & 97 & 0.81 $\pm$ 0.10 \\
900 - 1000 & 1285 & 495 & 474 & 78 & 0.88 $\pm$ 0.11 \\
1000 - 1200 & 1426 & 571 & 646 & 92 & 0.74 $\pm$ 0.10 \\
1200 - 2400 & 1685 & 521 & 614 & 74 & 0.73 $\pm$ 0.15 \\
\hline
Inclusive & 1266 & 3268 & 3542 & 563 & 0.76 $\pm$ 0.11 \\
\end{tabular}
\end{ruledtabular}
\end{table}

\subsection{\label{sbsec:cc1piErrs} Systematic errors}

  The systematic error on the neutrino flux measurment using the anti-neutrino-mode CC1$\pip$ sample comes from two sources that are treated as uncorrelated with each other: the uncertainty on the CC1$\pip$ cross section obtained from~\cite{CCpip} and the uncertainty in the background prediction.  The largest contribution to the uncertainty on the CC1$\pip$ cross section comes from the neutrino-mode flux uncertainty, which is the only systematic error associated with the cross-section measurement that is also independent of the measurement made here.  Because the other CC1$\pip$ uncertainties are treated as uncorrelated, a partial cancellation of errors is ignored in the present neutrino flux measurement.  This results in a slight overestimate of the neutrino flux uncertainty.  Both $\numu$ and $\numub$ background events in the sample are assigned 30\% uncertainties to conservatively recognize the model dependence of the sample composition.  The fractional uncertainty contributions to the flux measurement are presented in Table~\ref{tbl:ccpiFracErr}.

\begin{table}[h]
\caption{\label{tbl:ccpiFracErr} Fractional uncertainty (\%) contributions to the neutrino flux measurement in the CC1$\pi^{+}$ sample.  The $\numu$ uncertainty is dominated by the CC1$\pip$ cross-section error.}
\begin{ruledtabular}
\begin{tabular}{ccccc}
E$_{\nu}^{\mathrm{QE}}$ Range & \multirow{2}{*}{Statistical} & \multirow{2}{*}{$\numub$} & \multirow{2}{*}{$\numu$} & Total\\
(MeV) & & & & Fractional Error \\
\hline
600 - 700 & 6 & 9 & 11 & 15 \\
700 - 800 & 5 & 7 & 10 & 13 \\
800 - 900 & 5 & 6 & 10 &  13 \\
900 - 1000 & 5 & 6 & 10 & 13 \\
1000 - 1200 & 5 & 6 & 11 & 13 \\
1200 - 2400 & 5 & 5 & 19 & 20 \\
\hline
Inclusive & 2 & 6 & 13 & 14 \\
\end{tabular}
\end{ruledtabular}
\end{table}

\section{\label{QE} Measuring the Neutrino Flux Through Muon Angular Distributions in the CCQE Sample}

\subsection{\label{sec:QEsam} The CCQE sample}


The CCQE interaction is the dominant channel in MiniBooNE's energy range.  CCQE events typically have two subevents, attributed to the primary muon and the associated decay positron:

\beq
\begin{array}{cccl}
1: & \numub + p  & \to &  \mup + n \\ 
2: &            &     &  \hookrightarrow \elp + \nue + \numub. 
\end{array} 
\eeq

\noindent The CCQE sample is therefore similar in formation to the CC1$\pip$ sample with one major divergence: a requirement of two subevents instead of three.  As shown in Table~\ref{tbl:Effs}, the CCQE selection cuts closely follow those motivated in Section~\ref{sbsec:3se}, with a few exceptions appropriate to the inclusion of a single Michel electron.  The Michel tank hit and veto PMT hit cuts apply to the second subevent only now (Cuts 5 and 6, respectively), and the muon endpoint-electron vertex cut in Cut 7 is tightened to 100~cm in light of larger backgrounds. The selection cuts outlined here are identical to those employed in a previous shape-only extraction of CCQE model parameters~\cite{MB_PRL} and closely follow those used in the absolute measurement of the $\numu$ induced CCQE cross section~\cite{qePRD}, with only minor differences that result in approximately the same sample efficiency and purity.  

\begin{table}
\caption{\label{tbl:Effs} Summary of selection cuts with efficiencies in the CCQE sample. ``Purity" refers to $\numub$ CCQE only, and purity and efficiency numbers are sequential.}
\begin{ruledtabular}
\begin{tabular}{cccc}
 \multirow{2}{*}{Cut \#} &  \multirow{2}{*}{Description} & Efficiency & Purity \\ 
& & (\%) & (\%) \\
\hline
\vspace{1 mm}
0 & No cuts & 100 & 32 \\
\vspace{1 mm}
1 & Two subevents & 49 & 41 \\
 \multirow{2}{*}{\,\,2} & 1st subevent in event time window & \multirow{2}{*}{47} & \multirow{2}{*}{42} \\
\vspace{1 mm}
 & 4000 \textless\, T(ns) \textless\, 7000 & \\
 \multirow{2}{*}{\,\,3} & 1st subevent: reconstructed & \multirow{2}{*}{38} & \multirow{2}{*}{43}  \\
\vspace{1 mm}
 & vertex \textless\, 500~cm from tank center & & \\
\vspace{1 mm}
4 & 1st subevent: tank hits \textgreater\, 200 & 35 & 45 \\
\vspace{1 mm}
5 & 2nd subevent: tank hits \textless\, 200 & 33 & 45 \\
\vspace{1 mm}
6 & Both subevents: veto hits \textless\, 6 & 29 & 49 \\
\multirow{3}{*}{\,\,7} & Distance between reconstructed & \multirow{3}{*}{25} & \multirow{3}{*}{54}\\
& end of 1st subevent and 2nd & & \\
& subevent vertex \textless\, 100 cm & & \\
\end{tabular}
\end{ruledtabular}
\end{table}

Despite the selection cuts, there are formidable backgrounds to the anti-neutrino-mode CCQE sample.  Prior to this analysis, simulation estimates the anti-neutrino-mode CCQE sample has a purity just above 50\% as shown in Table~\ref{tbl:bkgrnds}.  The major backgrounds include CC1$\pip$ and CC1$\pim$ events, which account for a total of $\sim$ 20\% of the sample, and the $\numu$ processes, predicted to be responsible for $\sim$ 30\% of the sample.  The 30\% predicted $\numu$ contamination is investigated and ultimately constrained in this analysis.  

\begin{table}
\caption{\label{tbl:bkgrnds} Predicted composition of the anti-neutrino-mode CCQE sample.}
\begin{ruledtabular}
\begin{tabular}{cc}
Channel & Contribution (\%) \\
\hline
$\numub p \to \mup n$ & 54 \\
$\numu n \to \mum p$ & 20 \\
$\numub N \to \mup \pim N$ (resonant) & 8 \\
$\numu N \to \mum \pip N$ (resonant) & 6 \\
$\numub A \to \mup \pim A$ (coherent) & 4 \\
$\numub N \to \mup \Lambda, \Sigma$ & 3 \\
$\numub p \to \mup \piz n$ & 2 \\
Other & 3\\
\hline
All $\numub$ & 71 \\
All $\numu$ & 29 \\
\end{tabular}
\end{ruledtabular}
\end{table}

A few additional modifications to the simulation are made to accommodate the backgrounds.  The largest non-CCQE background in the sample is single pion production which enters the sample due to nuclear effects, including $\mu^{-}$, $\pi^{-}$ capture and final-state interactions; however, in the case of anti-neutrino induced CC1$\pim$ scattering, due to $\pim$ nuclear capture almost 100\% of CC1$\pim$ events have only two subevents and are experimentally indistinguishable from CCQE.  This implies a direct background measurement of CC1$\pim$ events (analogous to what was done in Ref.~\cite{qePRD}) is impossible.  Therefore, though the CC1$\pip$ yield constraint made in Ref.~\cite{qePRD} is strictly appropriate to neutrino induced CC1$\pip$ events only, it is applied to both predicted CC1$\pip$ \emph{and} CC1$\pim$ background events in the CCQE sample.  

Many backgrounds to the CCQE sample peak in the most forward scattering region of the muon angular distribution with respect to the incoming neutrino beam.  This includes pion production and hydrogen CCQE scattering - while the latter is technically not a background, the proper handling of the difference in nuclear effects between bound and free targets is not straightforward.  Additionally, the forward scattering region is strongly correlated with low-Q$^{2}$ events, a problematic region both experimentally and theoretically~\cite{BenharLowQ2}. Such low Q$^{2}$ data are dominated by $\numub$ interactions, while the present analysis is principally interested in backwards scattering muons which is dominated by $\numu$.  For these reasons, events with cos~$\thetmu\,~\textgreater\,~0.91$ are not included in the fit to data, where $\thetmu$ is the outgoing muon angle relative to the incoming neutrino beam.

\subsection{\label{sbsec:QEevntReco} CCQE event reconstruction}

\begin{figure}[h]
\begin{center}
\includegraphics[scale=0.46]{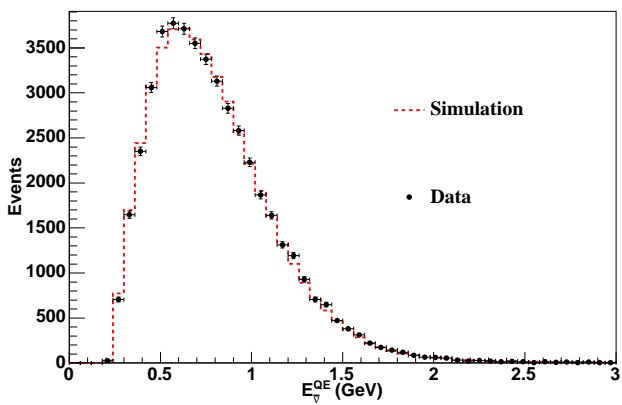}
\caption{(Color online) The reconstructed energy spectrum for simulation versus data in the anti-neutrino-mode CCQE sample. Simulation is normalized to data, and only statistical errors are shown.}
\label{fig:EnuQEsmear}
\end{center}
\end{figure}

Event reconstruction in the anti-neutrino-mode CCQE sample proceeds similar as in the CC1$\pip$ sample, described in Section~\ref{sbsec:ccpiEvntReco}.  As in the CC1$\pip$ reconstruction, measurement of muon kinematics from the primary interaction is solely responsible for recreating the incident neutrino energy.  No requirement is made on the ejected nucleon; this is an important distinction from the CCQE definitions used by other experiments~\cite{SciBooNE,NOMAD}, where a single proton track may be required in the case of neutrino-induced CCQE.  A similar energy reconstruction as described in Section~\ref{sbsec:ccpiEvntReco} is implemented, but in this sample a $\numub$ probe is assumed:

\begin{center}
\begin{equation}
\label{eqn:EnuQE}
E_{\bar{\nu}}^{\mathrm{QE}}= \frac{2 \left(M_{p} - E_{B}\right) E_{\mu} - \left( E^{2}_{B} - 2 M_{p} E_{B} + m^{2}_{\mu} + \Delta M^{2} \right)}{2\left[ \left(M_{p} - E_{B} \right) - E_{\mu} + p_{\mu}\,\text{cos}\,\theta_{\mu}\right]}\end{equation}
\par\end{center}

\noindent where the same definitions from Equation~\ref{eqn:EnuPi} apply and $\Delta M^{2}~=~M_{p}^{2}~-~M_{n}^{2}$, where $M_{n}$ is the neutron mass.  Figure~\ref{fig:EnuQEsmear} presents the reconstructed energy distributions in simulation and data in the CCQE sample.  CCQE scattering with free protons in hydrogen are indistinguishable from those on bound protons in carbon, so all events in data and simulation are reconstructed using the carbon scattering assumption implicit in Equation~\ref{eqn:EnuQE}.

\subsection{\label{sbsec:QEmeas} Neutrino flux measurement using CCQE}

\begin{figure}[h]
\begin{center}
\includegraphics[scale=0.46]{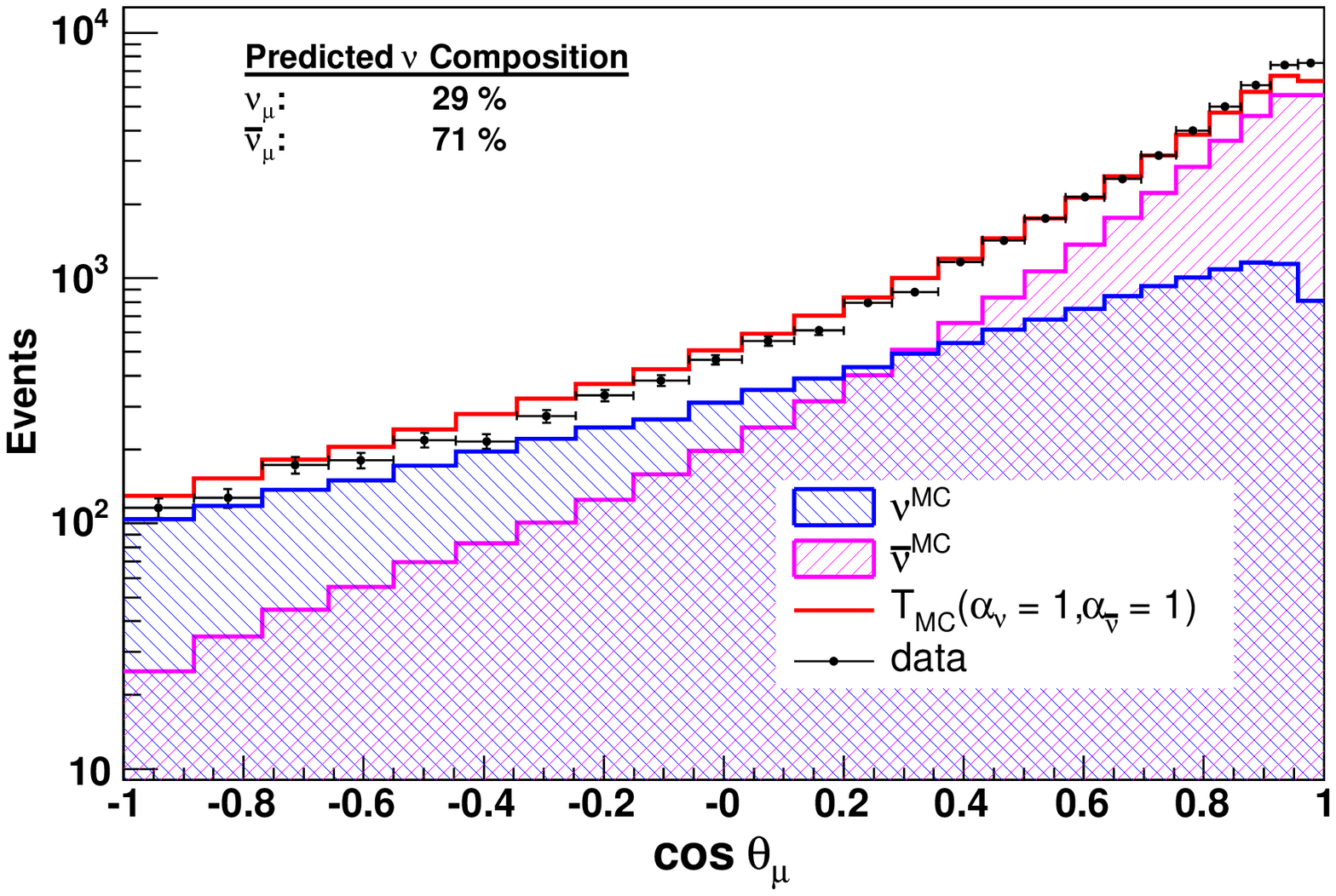}
\caption{(Color online) The cos~$\theta_{\mu}$ distribution of the CCQE sample by neutrino type before fitting.  As printed on the figure, 29\% of the sample is predicted to be induced by neutrinos.  The Monte Carlo sample has been normalized to 5.66 $\times$ 10$^{20}$ protons on target.}
\label{fig:QEdists}
\end{center}
\end{figure}

Neutrino and anti-neutrino CCQE cross sections differ exclusively by an axial-vector interference term that amplifies $\nu$ scattering while suppressing $\bar{\nu}$ events.  A particularly clean way to exploit this cross section difference is to fit the angular distribution of the primary muon.  The contribution from $\numub$ is suppressed in the backward scattering region.  Figure~\ref{fig:QEdists} shows the predicted $\numu$ and $\numub$ contributions to the cosine of the outgoing muon angle.

\indent To measure the neutrino content in the anti-neutrino mode beam, the Monte Carlo (MC) sample is separated into two cos~$\theta_{\mu}$ templates, one arising from all $\numu$ interactions and the other from $\numub$, regardless of interaction channel and nuclear target.  A linear combination of these two templates is then formed,

\begin{center}
\begin{equation}
T_{MC}(\alpha_{\nu},\alpha_{\bar{\nu}}) \equiv \alpha_{\nu}\,\nu^{MC} + \alpha_{\bar{\nu}}\,\bar{\nu}^{MC}\end{equation}
\label{eqn:fitfcn}
\par\end{center}

\noindent where $T_{MC}$ is the total predicted cos~$\theta_{\mu}$ distribution to be compared to data, $\alpha_{\nu}$ and $\alpha_{\bar{\nu}}$ are neutrino and anti-neutrino rate scales, and $\nu^{MC}$ and $\bar{\nu}^{MC}$ are the MC neutrino and anti-neutrino scattering angular predictions, respectively.  The modified simulation sample is compared to data by forming a goodness-of-fit $\chi^{2}$ test as a function of the rate scales: 


\begin{equation}
\begin{array}{l}
\chi^{2} = \displaystyle\sum_{i,j} \left( T_{MC}(\alpha_{\nu},\alpha_{\bar{\nu}})_{i} - d_{i}\right) M_{ij}^{-1} \left( T_{MC}(\alpha_{\nu},\alpha_{\bar{\nu}})_{j} - d_{j} \right)\\
\end{array} 
\label{eqn:chiSq}
\end{equation}


\noindent where $i$ and $j$ label bins of cos $\thetmu$, $d$ is data and $M$ is the symmetric error matrix given in Equation~\ref{eqn:ErrMat}.  The error matrix is used to propagate correlated uncertainties on parameters and processes to the quantities reported in the analysis.  It is made by first forming weights corresponding to simulation excursions set by Gaussian variations of parameters within their associated error.  The difference of these weighted events from the simulated central value forms the error matrix,

\begin{equation}
\begin{array}{l}
M_{ij} = \frac{1}{K}\sum\limits_{s = 1}^{K} (N_i^{s} - N_i^{CV}) \times (N_j^{s} - N_j^{CV}). \\   
\end{array} 
\label{eqn:ErrMat}
\end{equation}

\noindent Here $K$ simulation excursions are used ($K$ = 100 in this analysis), $N^{s}$ is the re-weighted number of entries corresponding to the $s^{th}$ simulation set and $N^{CV}$ represents the MiniBooNE simulation central value.  This technique is further described in Ref.~\cite{brNIM}. Bin-by-bin cos $\thetmu$ correlations between $\numu$ and $\numub$ are also treated.  The specific systematic errors are discussed in the next section.




The fit is performed analytically in three bins of reconstructed energy and also in an inclusive energy sample.  Results including statistical and systematic uncertainties are presented in Table~\ref{tbl:results}.  The fits to data are shown in Appendix~\ref{apndx:QEplats}, where Figure~\ref{fig:fits} contains both the fitted distributions and the fractional differences between the simulation and data before and after the fits.  The adjusted contributions of $\numu$ and $\numub$ to the CCQE sample are compared to the prediction in Table~\ref{tbl:fracs}.

\begin{table*}
\caption{\label{tbl:results} Fit results in three energy bins and an inclusive sample.  The results are consistent with an over-prediction of the $\numu$ contamination of the MiniBooNE anti-neutrino-mode CCQE sample.}
\begin{ruledtabular}
\begin{tabular}{ccccccc}
E$_{\bar{\nu}}^{\mathrm{QE}}$ Range & Mean Generated & Events & \multirow{2}{*}{$\alpha_{\nu}$ fit} & \multirow{2}{*}{$\alpha_{\bar{{\nu}}}$ fit} & $\rho_{\alpha_{\nu}-\alpha_{\bar{\nu}}}$ fit &  $\chi^{2}$  \\
(MeV) & E$_{\nu}$ (MeV) & in Data & & & correlation & (DOF = 21) \\
\hline
\textless\, 600 & 675 & 15242 & 0.65 $\pm$ 0.22 & 0.98 $\pm$ 0.18 & 0.33 & 13 \\
600 - 900 & 897 & 16598 & 0.61 $\pm$ 0.20 & 1.05 $\pm$ 0.19 & 0.49 & 21 \\
\textgreater\, 900 & 1277 & 15626 & 0.64 $\pm$ 0.20 & 1.18 $\pm$ 0.21 & 0.45 & 7 \\
\hline
Inclusive & 950 & 47466 & 0.65 $\pm$ 0.23 & 1.00 $\pm$ 0.22 & 0.25 & 16 \\
\end{tabular}
\end{ruledtabular}
\end{table*}

\begin{table}
\caption{\label{tbl:fracs} Fractional composition of the anti-neutrino-mode CCQE sample before and after angular fits.}
\begin{ruledtabular}
\begin{tabular}{ccccc}
E$_{\bar{\nu}}^{\mathrm{QE}}$ Range & \multicolumn{2}{c}{Before Fit (\%)} & \multicolumn{2}{c}{After Fit (\%)} \\
(MeV) & $\numu$ & $\numub$ & $\numu$ & $\numub$ \\
\hline
\textless\, 600                       & 25 & 75 & 18 $\pm$ 6 & 82 $\pm$ 16 \\
600 - 900 & 26 & 74 & 17 $\pm$ 6 & 83 $\pm$ 15 \\
\textgreater\, 900                    & 35 & 65 & 23 $\pm$ 7 & 77 $\pm$ 15  \\
\hline
Inclusive & 29 & 71 & 21 $\pm$ 8 & 79 $\pm$ 18 \\
\end{tabular}
\end{ruledtabular}
\end{table}

The $\chi^{2}$ value for the angular fit in the reconstructed energy range $E_{\nu}^{\mathrm{QE}}\,\textgreater\,900$ MeV is unusually low at $\chi^{2}$ = 7 for 21 degrees of freedom.  This is believed due simply to chance, as the statistical error only fit agrees with the data exceptionally well within the error, returning $\chi^{2}$ = 13 for 21 degrees of freedom. 

As the $\numu$ angular template has been corrected for the observed cross section per Ref.~\cite{qePRD}, $\alpha_{\nu}$ may be interpreted as a flux scale factor, and significant deviations from unity would imply a flux mismodeling.  Consistent with the results reported in Section~\ref{sbsec:ccpiWSmeas}, fits in the anti-neutrino-mode CCQE sample indicate the true neutrino flux to be somewhat lower than the simulation predicts.  Over all reconstructed energies, the neutrino flux component of the anti-neutrino-mode beam should be scaled by 0.65 to match the observed data.  Fits in individual reconstructed energy bins show that the neutrino flux component shape is well-modeled.  Finding the calibration on the neutrino flux component inconsistent with unity is not surprising, as the neutrino parent pions originate primarily in a poorly constrained production region (cf. Figure~\ref{fig:pionAngles}).  The rate scale $\alpha_{\bar{\nu}}$ is ambiguous in interpretation, as the cross section is yet unmeasured.  

\indent Care must be taken when comparing these results to the $\mup$/$\mum$ yield numbers reported in the MiniBooNE $\numub \to \nueb$ oscillation analysis~\cite{nubOsc1,nubOsc2}, since the interaction prediction is different.  In the oscillation analysis the cross section parameters measured in Ref.~\cite{MB_PRL} are employed, which includes $M_{A}^{\mathrm{eff}} = 1.23 (1.13)$ GeV for bound (free) nucleon CCQE scattering and $\kappa = 1.022$.  When the muon angular fit technique described in this section is repeated with this prediction, yield rates of $\alpha_{\nu} = 0.99 \pm 0.23$ and $\alpha_{\bar{\nu}} = 1.20 \pm 0.23$ are found, as reported in Ref.~\cite{nubOsc2}.  With this alternate CCQE scattering model, the angular fit over all reconstructed energies reports a neutrino contamination in the sample of 23 $\pm$ 6\%, consistent with the 21 $\pm$ 8\% contamination found with the scattering assumptions described in Section~\ref{sbsec:QExsec}.

\indent The results from this technique depend on knowing the angular distributions of neutrino and anti-neutrino CCQE interactions in the detector.  While the procedure relies on exploiting the effect of the interference term in the CCQE cross section, the angular distributions may be somewhat altered by nuclear effects.  In this analysis the measured angular distribution of neutrino interactions on carbon~\cite{qePRD} is employed, but the measurement relies on the scattering model described in Section~\ref{sbsec:QExsec} to predict anti-neutrino interactions. This model does not include two body current effects which may be larger than previously expected~\cite{Martini1} and may introduce additional neutrino and anti-neutrino angular differences. Despite this inherent model dependence, the results present a demonstration of a technique aimed at informing future experiments looking to separately constrain neutrino and anti-neutrino events in an unmagnetized environment.  By that time, the effect of additional nuclear processes on the angular dependence of anti-neutrino CCQE scattering should be better known.

\subsection{\label{sbsec:QEsystErrs} Systematic errors}

As the present analysis directly measures the neutrino component in the anti-neutrino-mode beam, systematic errors relating to beam geometry and meson production at the target are not considered.  The remaining systematic errors include those arising from detector modeling, the single pion production background, and the cross section parameters in the underlying model. Contributions propagated from these errors to the uncertainty on the parameter $\alpha_{\nu}$ in the inclusive energy sample are given in Table~\ref{tbl:systErrsQE}.  

\begin{table}
\caption{\label{tbl:systErrsQE} Summary of systematic error contribution to the scale parameter $\alpha_{\nu}$ in the inclusive energy fit.  Individual error contributions are found for the $i$th systematic error by first repeating the fits with only independent systematics considered.  The fractional error contributions are then found by $\sqrt{(\De \alpha_{\nu}/\alpha_{\nu})^{2}_{syst_{i} + stat} - (\De \alpha_{\nu}/\alpha_{\nu})^{2}_{stat}}$, where $\De \alpha_{\nu}$ is the one-sigma error reported in Table~\ref{tbl:results}.  The statistical error is found by considering the second term only.  This method does not account for small changes in the $\alpha_{\nu}$ best fit parameter between the fits considering various errors, and so the individual fractional errors do not add in quadrature to produce the total fractional error reported in Table~\ref{tbl:results} and in the final column.}
\begin{ruledtabular}
\begin{tabular}{cc}
Source of Error & Fractional Uncertainty (\%) \\
\hline
Statistical & 8 \\
Detector Modeling & 11 \\
CC1$\pip$ Constraint & 4 \\
Cross Section & 26 \\
\hline
Total Fractional Error & 35 \\
\end{tabular}
\end{ruledtabular}
\end{table}

\indent Apart from final-state interaction uncertainties leading to errors on the cross section, the error on the CC1$\pip$ background contributes to the systematic error through the error labeled ``CC1$\pip$ Constraint" in Table~\ref{tbl:systErrsQE}.  This measurement uncertainty is based on a $Q^{2}$-dependent shape-only scale factor to improve data-simulation agreement in the neutrino-mode CC1$\pip$ sample~\cite{prlDisap}.  The cross section (both CCQE and CC1$\pip$) uncertainty is dominant in these fits and warrants further discussion.  Table~\ref{tbl:xsecErr} offers a breakdown of cross section parameters and associated errors.  The error on carbon $M_{A}^{\mathrm{eff}}$ has been reduced from that reported in Ref.~\cite{qePRD} to avoid double-counting MiniBooNE systematic errors applicable to both the measurement of $M_{A}^{\mathrm{eff}}$ and the measurement reported here.  The 26\% uncertainty due to cross-section errors reported in Table~\ref{tbl:systErrsQE} can be expanded as the quadrature sum of 16\% from the 10\% normalization errors on $\numub$ and CCQE processes, 14\% from the error on M$_{A}$ and $\kappa$, and 15\% from the remaining.

\begin{table}[h]
\caption{\label{tbl:xsecErr} Summary of cross-section errors used in this analysis.  The bottom portion presents fractional uncertainties assigned to processes \emph{in addition} to parameter errors.  Errors given on pion absorption and charge exchange are relevant to pion propagation in the detector medium.}
\begin{ruledtabular}
\begin{tabular}{cc}
Parameter & Value with Error \\
\hline
$M_{A}^{\mathrm{eff}}$ carbon target & 1.35 $\pm$ 0.07~GeV \\
$M_{A}^{\mathrm{eff}}$ hydrogen target & 1.03 $\pm$ 0.02~GeV \\
$\kappa$ & 1.007 $\pm$ 0.005 \\
$E_{B}$ & 34 $\pm$ 9~MeV \\
$\Delta$s & 0.0 $\pm$ 0.1 \\
M$_{A}^{1\pi}$ & 1.10 $\pm$ 0.28~GeV \\
M$_{A}^{\mathrm{multi-\pi}}$ & 1.30 $\pm$ 0.52~GeV \\
p$_{F}$ & 220 $\pm$ 30~MeV \\
\hline
\multirow{2}{*}{Process} & Fractional \\
& Uncertainty (\%) \\
\hline
$\pip$ Charge Exchange & 50 \\
$\pip$ Absorption & 35 \\
CCQE $\sigma$ Normalization & 10 \\
All $\numub$ $\sigma$ Normalization & 10 \\
$\De$ + N $\to$ N + N & 100 \\
\end{tabular}
\end{ruledtabular}
\end{table}

As the main contributions to the dominant cross section systematic error apply to both $\numu$ and $\numub$ scattering, $\alpha_{\nu}$ and $\alpha_{\bar{\nu}}$ are positively correlated as reported in Table~\ref{tbl:results}.

\section{\label{sec:comp} Result Comparison}

\begin{figure}[h]
\begin{center}
\includegraphics[scale=0.46]{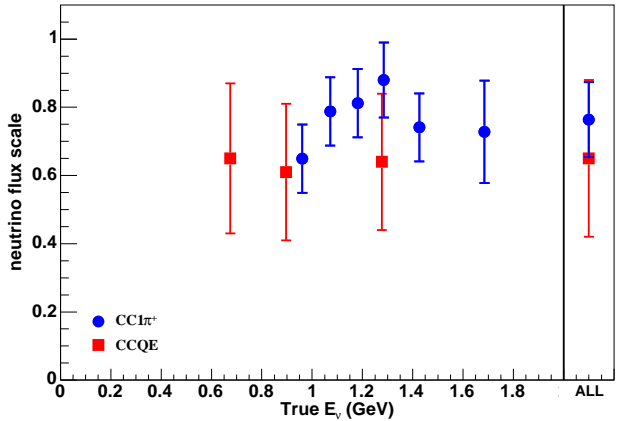}
\caption{(Color online) Summary of the neutrino flux constraint in the anti-neutrino-mode beam from the CC1$\pip$ (Section~\ref{sec:CC1pi}) and CCQE (Section~\ref{QE}) measurements.}
\label{fig:moneyPlat}
\end{center}
\end{figure}

Including all reconstructed energies in the CC1$\pip$ sample, a neutrino flux component scale of 0.76 $\pm$ 0.11 is found, while the CCQE analysis yields 0.65 $\pm$ 0.23.  The measurements are compatible and complementary as each analysis includes energy regions not covered by the other as shown in Fig.~\ref{fig:moneyPlat}.  The results indicate the simulated neutrino component of the anti-neutrino-mode flux is overestimated by $\sim$ 30\%.  These flux measurements constrain the very forward $\pip$ created at the target, where an external data constraint is not available.  Results from both methods are summarized in Figure~\ref{fig:moneyPlat}, where measurements are placed at the mean of the generated energy distribution for each reconstructed energy sample.
\vspace{10mm}

\section{\label{sec:imps} Implications for other experiments}

The techniques applied here could also aid future neutrino experiments
  that will test for CP violation in the lepton sector using large unmagnetized
  detectors. This includes experiments such as NO$\nu$A~\cite{nova}, T2K~\cite{t2k}, LBNE~\cite{lbne},
  LAGUNA~\cite{memphys}, and Hyper-K~\cite{hyperk}.  A magnetized near detector can provide a powerful constraint on the neutrino flux and provide precise cross sections.  However, a measurement of the neutrino rate at the far detector can still be very useful given that the the beam spreads from an extended source and oscillates while traveling between the detectors.
  
  Additional techniques could offer potentially
  helpful constraints on the neutrino component in an anti-neutrino-mode beam.
  This includes taking advantage of the effective lifetime difference between
  $\mu^-$/$\mu^+$ due to $\mu^-$ capture in a nuclear environment. Fitting the lifetime
  distributions or measuring how often a decay electron is produced could supply
  constraints that are especially useful as they are independent of the underlying
  neutrino interaction cross sections. Also, selection of CCQE interactions with
  and without a proton in the final state may afford additional neutrino versus
  anti-neutrino tagging capabilities~\cite{pTag1,pTag2}.

\section{\label{sec:conc} Conclusions}

Two analyses are presented to measure the neutrino flux in the MiniBooNE anti-neutrino-mode beam.  The two measurements have a common dependence on the neutrino flux in the neutrino-mode beam~\cite{mbFlux} that MiniBooNE obtained from HARP hadroproduction data. At present the CCQE angular distribution method is largely limited by uncertainties in the cross sections, especially the anti-neutrino cross section which MiniBooNE is in the process of measuring, while the uncertainty of the CC1$\pi$ method is dominated by the neutrino-mode flux uncertainty.

Using two event samples dominated by independent physics processes, compatible and complementary results are found.  The results from both analyses indicate the prediction of the neutrino flux component of the anti-neutrino beam is over-estimated - the CC1$\pip$ analysis (Section~\ref{sec:CC1pi}) indicate the predicted $\numu$ flux should be scaled by $0.76 \pm 0.11$, while the CCQE angular fits (Section~\ref{QE}) yield $0.65 \pm 0.23$.  Results from repeating the analyses in bins of reconstructed neutrino and anti-neutrino energy indicate that the predicted flux spectrum shape is well modeled.  The results from fitting the muon angular distributions in the CCQE sample has already been employed in the MiniBooNE oscillation analysis~\cite{nubOsc1,nubOsc2}, while the CC1$\pip$-based measurement will likely be more valuable to MiniBooNE anti-neutrino cross section extractions, as it is much less model dependent and carries comparatively smaller uncertainty.  These types of analyses, along with others discussed in Section~\ref{sec:imps} may be of use to present and future precision neutrino experiments testing CP violation with neutrino and anti-neutrino beams.

We wish to acknowledge the support of Fermilab, the National Science Foundation, and the Department of Energy in the construction, operation, and data analysis of the MiniBooNE experiment.


\newpage
\clearpage

\appendix
\section{\label{apndx:QEplats} CCQE angular fit details}

\begin{figure*}[h]
\begin{center}$
\begin{array}{cc}
\includegraphics[scale=0.90]{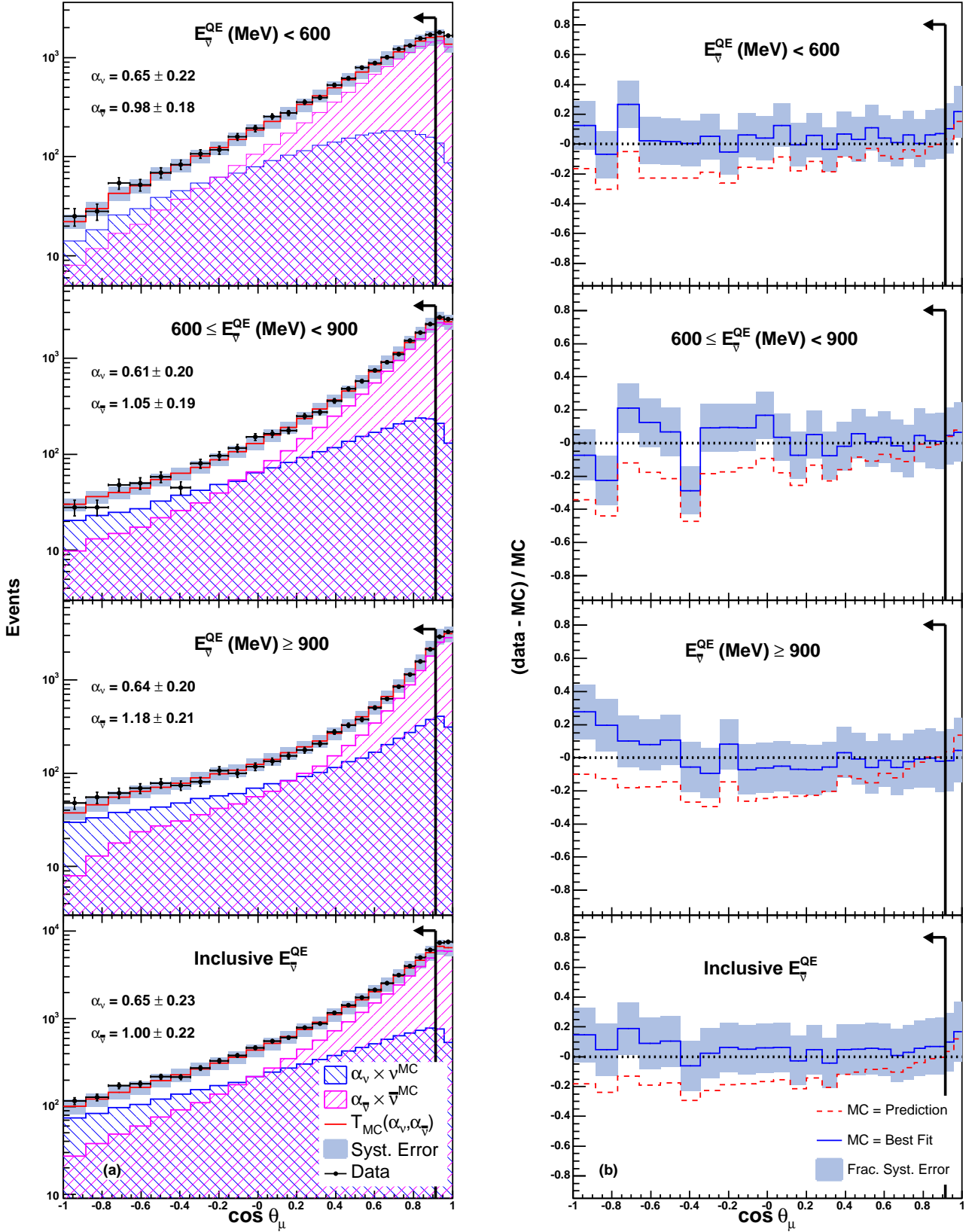} &
\end{array}$
\end{center}
\caption{\label{fig:fits} (Color online) Results of the muon angular fits to the CCQE data described in Section~\ref{QE}.  Shown are (a) the fits and (b) fractional differences (data - simulation)~/~simulation for both the unmodified prediction and the best fit.  Along with an inclusive sample, three reconstructed energy bins are considered.  The before-fit simulation is absolutely normalized to 5.66 $\times\, 10^{20}$ protons on target.  Only events with cos $\thetmu$~\textless~0.91 are considered.}
\end{figure*}

This appendix presents details on the CCQE angular fit results described in Section~\ref{QE}. The fits to data are plotted in Figure~\ref{fig:fits}.


\end{document}